\documentclass[aps,pra]{revtex4}
\usepackage{graphicx}

\begin{document}

\title{The security proof of the ping-pong protocol is wrong
\thanks{Email: zjzhang@ahu.edu.cn}}
\author{Zhan-jun Zhang\\
{\normalsize School of Physics \& Material Science, Anhui University, Hefei 230039, Anhui, China} \\
Email address: zjzhang@ahu.edu.cn}

\date{\today}
\maketitle

In 2002, Bostrom and Felbinger presented a novel quantum secure
direct communication protocol (named often as ping-pong protocol)
[1] with a security proof in the case of ideal quantum channel. In
this paper, we will show that the security proof is wrong in the
strict sense.

To prove the security of their protocol, Bostrom and Felbinger
assume that Eve adds an ancilla in the state $|\chi\rangle$ and
performs a unitary operation $\hat{E}$ on the composite system
consisting of the travel qubit and the ancilla. Then they worked
out the von-Neumann entropy $S$ of the state of the travel qubit
after Eve's attack operation and after Alice's encoding operation.
Von-Neumann entropy $S$ is taken as the maximal amount $I_0$ of
classical information that can be extracted from a state. After
some deductions, they got function $I_0(d)$, where $d$ is the
detection probability for Eve's attack. By analyzing $I_0(d)$,
they conclude that {\it any effective eavesdropping attack can be
detected}. However, it should be emphasized that in Ref.[1] the
$I_0$ is extracted from the travel-qubit state. This is clearly
stated by Bostrom and Felbinger themselves nearby the equation 8
in Ref.[1]. Incidentally, if the equations 7 and 8 in Ref.[1]
denote the composite-system state, then the calculation of $S$ is
wrong according to Ref.[2]. As will also lead to the fact that the
security proof is wrong.

About Bostrom and Felbinger's proof, there exists a very serious
question: Is it reasonable to assume in priori that Eve extracts
useful information {\it only from the travel-qubit state} and
disregards the ancilla state and the composite-system state? Only
at a glance, one will answer {\it no}, because if the ancilla
state and the composite-system state are completely useless, then
it is completely unnecessary for Eve to introduce an ancilla. Now
let us extensively analyze the above question. Relative to the
proof, especially to the analysis of $I_0(d)$ in the proof, the
question can be transformed into another one. That is, whether
$I_{0a}$ and $I_{0c}$ are always not greater than $I_{0t}$ for any
$d$, where $I_{0t}$, $I_{0a}$ and $I_{0c}$ denote the maximal
amounts of classical information extracted from the travel-qubit
state, the ancilla state and the composite-system state,
respectively. If the answer is {\it yes}, then the priori
assumption is of course reasonable, because the analysis of
$I_{0t}(d)$ is enough for security proof. Otherwise, Bostrom and
Felbinger's proof will be quite questionable. This is because
$I_{0a}(d)$ and $I_{0c}(d)$ are not obtained and consequently one
does not know whether Eve can extract some useful information from
the ancilla state or the composite-system state and meanwhile
faces zero detection probability. In this case, Bostrom and
Felbinger's security proof is insufficient to support their
conclusion that {\it any effective eavesdropping attack can be
detected}. Of course, the proof is unconvinced and one can say the
security proof is wrong. Hence, now the key question is that
whether $I_{0a}$ and $I_{0c}$ are always not greater than $I_{0t}$
for any $d$. In fact, the key question can be easily answered via
a simple example. Suppose Bob sends $|0\rangle$ and the ancilla
Eve adds is a qubit in the state
$|\chi\rangle=\frac{\sqrt{2}}{2}(|0\rangle+|1\rangle)$. Then the
state of the composite system consisting of the travel qubit and
the auxiliary qubit is
$|\xi\rangle=\frac{\sqrt{2}}{2}(|00\rangle+|01\rangle)$. Assume
that the unitary operation $\hat{E}$ Eve performs on the composite
system is $
%\begin{eqnarray}
\hat{E}=
\frac{\sqrt{2}}{2}(|00\rangle\langle00|-|00\rangle\langle10|+
|01\rangle\langle01|-|01\rangle\langle11|+|10\rangle\langle00|+|10\rangle\langle10|
+|11\rangle\langle01|+|11\rangle\langle11|). $
%\end{eqnarray}
In this case, after Eve's attack operation, the eavesdropping
detection probability $d$ is 1/2. Then after Alice's encoding
operation, $I_{0t}=1$ and $I_{0c}=2$ can be easily worked out.
Apparently, these two values satisfy Ref.[2]'s conclusion (i.e.,
{\it the completely mixed density operator in a $n$-dimensional
space has entropy $\log_2n$}). Since $I_{0c}$ is obviously greater
than $I_{0t}$ in the case of $d=1/2$,  of course, the proof's
priori requirement that $I_{0a}$ and $I_{0c}$ are always not
greater than $I_{0t}$ is denied. Hence, whether Eve can extract
some useful information from the composite-system state in the
case of $d=0$ becomes an unsolved question. Accordingly, Bostrom
and Felbinger's conclusion that {\it any effective eavesdropping
attack can be detected} is cursory. In the strict sense, their
so-called security proof of ping-pong protocol in Ref.[1] is wrong
for it does not indeed indicate that the ping-pong can not be
eavesdropped in an ideal quantum channel. Incidentally, the recent
work[3] has already revealed that the ping-pong protocol can be
eavesdropped even in an ideal quantum channel, as strongly
supports the conclusion of this paper.

This work is supported by the National Natural Science Foundation
of China under Grant No.10304022, the science-technology fund of
Anhui province for outstanding youth under Grant No.06042087, the
general fund of the educational committee of Anhui province under
Grant No.2006KJ260B, and the key fund of the ministry of education of China
under Grant No.206063.\\

\noindent {\bf References}

\noindent[1] K. Bostrom and T. Felbinger, Phys. Rev. Lett. {\bf
89}, 187902 (2002).

\noindent[2] M. A. Nielsen and I. L. Chuang, Quantum Computation
and Quantum Information,(Cambridge University Press, Cambridge,
2000) p510.

\noindent[3] Z. J. Zhang, Y. Li and Z. X. Man, Phys. Lett. A {\bf
341}, 385 (2005).

\enddocument